\documentclass{aa}
\input psfig.tex
\usepackage{graphicx}
\usepackage{natbib}
\usepackage{array}
\bibpunct{(}{)}{;}{a}{}{,}
        \newcommand{\Hbeta}{\mbox{${\rm  H_{\beta}}$}}
        
        \newcommand{\MFe}{\mbox{${\rm \langle Fe \rangle} $}}
        \newcommand{\mgii}{\mbox{${\rm Mg_{2}} $}}
        
        \newcommand{\mgb}{\mbox{${\rm Mg_{b}} $}}

	\newcommand{\FeH}{\mbox{{\rm [Fe/H]}}}
	
	\newcommand{\MgsFe}{\mbox{{\rm [Mg/Fe]}}}
	\newcommand{\OsFe}{\mbox{{\rm [O/Fe]}}}
	\newcommand{\alfsFe}{\mbox{{\rm [$\alpha$/Fe]}}}
        \newcommand{\alfa}{\mbox{$\alpha$-elements}}
        \newcommand{\alfe}{\mbox{$\alpha$-enhanced}}

	\newcommand{\Xsun}{\mbox{${\rm X_{\odot}}$}}
	
	\newcommand{\Zsun}{\mbox{${\rm Z_{\odot}}$}}

        \def\smallskip{\vskip 4pt}

\begin{document}

\title{Enhancement of $\alpha$-Elements in Dynamical Models of Elliptical 
         Galaxies}
%\subtitle{}
     \author{ R. Tantalo\inst{1}
        \and
	      C. Chiosi\inst{1,2}
     }
     \offprints{R. Tantalo }
     \institute{$^1$ Department of Astronomy, University of Padova,
                Vicolo dell'Osservatorio 2, 35122 Padova, Italy \\
 $^2$ Visiting Scientist, ESO, Karl-Schwarzschild-str. 2, D-85748, 
 Garching bei Muenchen, Germany \\
     \email{tantalo@pd.astro.it;  chiosi@pd.astro.it; cchiosi@eso.org}
     }
     \date{Received: November, 2001; accepted ***}

%=================== BEGIN ABSTRACT =====================%
\abstract{
In this paper we present a quantitative analysis of the chemical
abundances in "monolithic" N-body-Tree-SPH models of elliptical
galaxies to determine whether the ratio \alfsFe\ increases with the
galaxy mass as suggested by observational data.
\keywords{Elliptical Galaxies; Dynamical Models; Chemical Models; 
Abundances; Spectral Indices } }

\titlerunning{Enhancement of $\alpha$-elements in elliptical galaxies }
\authorrunning{R. Tantalo \&\ C. Chiosi}
\maketitle

%=================== BEGIN SECTION ====================%
\section{Introduction}

Twenty years after the seminal papers on the formation of elliptical
galaxies (EGs) by \citet{Larson75} and \citet{Toomre77} there is still
much disagreement among astronomical community on both the process of
formation and evolution of early-type galaxies.

Are EGs formed in a single collapse during a short epoch of activity
(monolithic scheme) or continuously by hierarchical mergers at similar
levels by several mergers (hierarchical scheme)?  Once created, is the
stellar population of EGs just passively evolving or do minor
merging/accretion events drastically change their characteristics
frequently? What is the influence of density environment?

In the {\it monolithic scenario} EGs are supposed to suffer dominant
star forming activity and consequent chemical enrichment at very early
epochs followed by quiescent evolution, marginal episodes of star
formation of both internal and external origin may occur under
suitable circumstances (e.g. interactions).

In the {\it hierarchical scenario} EGs are supposed to be formed by
mergers (one to several) of smaller units, each episode inducing star
formation and chemical enrichment.

Both scenarios are able to reproduce only part of the observed
properties of EGs.  For instance the ``monolithic scheme'', on which
current semi-analytical models of chemical and spectro-photometric
evolution of EGs (see below) are based, cannot account for the wide
morphological and kinematical variety of EGs, for instance the
presence of counter-rotating cores, small disks, recent stellar
activity
\citep{Schweizer90,Schweizer92,Longhetti98a,Longhetti98b,Longhetti20}.

In contrast, despite its success in modeling the dynamical structure
of EGs, the ``hierarchical scenario'' suffers a point of difficulty as
far as the chemical properties are concerned. Indeed, EGs often have
different ratios between the abundances of some chemical species. They
are in general more metal-rich and enhanced in \alfa\ with respect to
spiral galaxies. In the merging scenario this could be justified by an
additional chemical enrichment with strong enhancement in \alfa\ must
take place.  See the reviews by \citet{Chiosi00,Chiosi01}.

Recently \citet{Chiocar01} have investigated by means of N-Body,
Tree-SPH models whether the monolithic scheme of galaxy formation and
evolution may interpret the observational properties of EGs.  They
have studied in particular how the history of star formation depends
on the initial conditions (density) and/or mass of the proto-galaxy
showing that the monolithic scheme can indeed reproduce many
fundamental properties of EGs, such as colors, mass to light ratios,
the color magnitude relation (CMR), and the so-called Fundamental
Plane \citep[see][ for all other details]{Chiocar01}. 

In this study we intend to examine whether the same models would also
predict chemical properties compatible with the observational data. In
particular we estimate their degree of enhancement in
$\alpha$-elements and check whether they can account for the
\alfsFe-mass relation of EGs (see below).

The plan of the paper is as follows. In Section 1 we discuss the
internal contradiction between the kind of constraint imposed by the
CMR and the \alfsFe-mass relationship. In Section 3 we present the
main results from the dynamical models, outline the companion chemical
models, and report on the results for the chemical abundances and
abundance ratios. In Section 4 we present an {\it ad hoc} experiment
aimed at pinning down the main parameter driving the level of \alfe\
in model galaxies. In Section 5 we summarize a few basic relations
defining the enhancement factor of a single stellar population (SSP)
of assigned chemical composition [X,Y,Z], present a simple algorithm
to calculate the spectral indices $\rm H_{\beta}$, $\rm Mg_2$, \MFe\
and $\rm Mg_b$, apply it to our model galaxies, and derive their mean
indices. Finally, these latter are compared with the observational
data paying particular attention to the $\rm Mg_2$ vs $\sigma$
(velocity dispersion) relation and the $\rm H_{\beta}$ vs $\rm Mg_2$
and $\rm H_{\beta}$ vs $[MgFe]$ planes.  In Section 6 we draw some
concluding remarks.

%=================== BEGIN SECTION ====================%
\section{A long lasting contradiction }

EGs show important properties that are used as key constraints to
theoretical models aimed at understanding the formation mechanism and
the past evolutionary history. Chief among these are the CMR and the
enhancement in \alfa\ suspected to exist in the brightest (most
massive) ellipticals.

%=================== BEGIN SUBSECTION ====================%
\subsection{The Color-Magnitude Relation}

The colors of EGs get redder at increasing luminosity and/or mass
\citep{Matthews71,Larson74,Bower92a,Bower92b}. The relation is tight
for cluster EGs \citep{Bower92a,Bower92b}, whereas it is more
disperse for nearby field EGs and those in small groups 
\citep{Schweizer92}.

The CMR is conventionally interpreted as a mass-metallicity sequence
\citep{Faber77,Dressler84b,Vader86b}: massive galaxies reach higher
mean metallicities than less massive ones. Starting from the original
suggestion by \citet{Larson74}, the mass-metallicity sequence is
attributed to the so-called supernova driven galactic wind (SDGW)
mechanism halting star formation at an age that depends on the galaxy
mass.  In short, the overall duration of the star forming activity
should be longer at increasing mass of the EGs ($\rm \Delta t_{SF}
\propto M_{G}$).

The tightness of the CMR is thought of to reflect the age of a galaxy
stellar population. The CMR of the Virgo and Coma clusters, where the
dispersion is very small, is compatible with most of the stars in EGs
being formed at redshift $z > 2$ which roughly corresponds to a
lookback time of about 10 Gyr for a Universe with $q_0=0.5$ and
$H_0=70$ km s$^{-1}$ Mpc$^{-1}$. The absolute value of the age is
less of a problem here. Nevertheless, secondary bursts of star
formation at $z=0.5$, forming about 10\% of the stellar mass could be
also accommodated \citep{Bower92a,Bower92b}. The possibility of
secondary activity has also been invoked by \citet{Kuntschner98a} and
\citet{Kuntschner00} on a different ground. In contrast the more
dispersed CMR of the field and small group EGs is compatible with ages
of the component stars spread over several billion years. This has
also been suggested using the line index diagnostic by
\citet{Gonzalez93} and \citet{Trager20a,Trager20b} with a tendency
for lower velocity dispersion (mass) systems to be younger \citep[see
also][]{Bressan96}.

%=================== BEGIN SUBSECTION ====================%
\subsection{The [$\alpha$/Fe]-Mass Relation}

Gradients in line strength indices $\rm Mg_2$ and \MFe\ (and others)
have been observed in EGs \citep{Worthey92,Gonzalez93,Davies93,Carollo93,
Carollo94a,Carollo94b,Balcells94,Fisher95,Fisher96}. The hint arises
that the indices are stronger toward the center. There is also the
controversial question whether the gradients in $\rm Mg_2$ and \MFe\
are the same. According to \citep{Carollo93,Carollo94a,Carollo94b} the
gradient in $\rm Mg_2$ is steeper than the gradient in \MFe, which is
interpreted as indicating that the abundances of \alfa\ (Mg, O, etc.)
with respect to iron are enhanced -- [$\alpha$/Fe]$>$0 -- in the
central regions. However, \citet{Kuntschner98} reaches the opposite 
conclusion and \citet{Davies2001} show for NGC4365 that the ratio [Mg/Fe] stays
constant within the data range.

Furthermore, limited to the nuclear regions, indices vary passing from
one galaxy to another \citep{Gonzalez93,Trager20a,Trager20b}. Looking
at the correlation between $\rm Mg_2$ and \MFe\ (or similar indices)
for the galaxies in the above quoted samples, $\rm Mg_2$ increases
faster that \MFe\, which is once more interpreted as due to
enhancement of \alfa\ in some galaxies. In addition to this, since the
classical paper by \citet{Burstein88}, the index $\rm Mg_2$ is known
to increase with the velocity dispersion (and hence mass and
luminosity) of the galaxy. Standing on this body of data the
conviction arose that the degree of enhancement in \alfa\ ought to
increase passing from dwarf to massive EGs \citep{Faber92,Worthey94,
Matteucci94,Matteucci97,Matteucci98}.

Understanding these properties rests on stellar nucleosynthesis and
star formation. In brief, all heavy elements observed in the
interstellar medium (ISM) are produced and ejected by stars, in
particular via supernova explosions. Furthermore, iron is mainly
produced by the long-lived Type Ia supernovae (accreting white dwarfs
in binary systems in the most popular scheme) whereas
\alfa\ are mainly generated by the short-lived Type II supernovae. 
Therefore, to enhance the \alfa\ there are several avenues:

\begin{itemize}
\item (i) A different ratio between Type Ia and Type II supernovae.
\item (ii) A different time scale of star formation.
\item (iii) A pre-enriched interstellar medium. 
\end{itemize}

The simplest and most widely accepted interpretation is the one based
on the different duration of the star forming period and the different
contribution to chemical enrichment by Type Ia and Type II
supernovae. In this context, it is worth recalling that the minimum
mean time scale for a mass accreting white dwarf in a binary system to
get the supernova stage is $\geq$0.5 Gyr. Therefore, the iron
contamination by Type Ia supernovae occurs later as compared to the
ones of Type II \citep{Greggio83}. It follows from this that
with the standard SDGW model and classical initial mass function
(IMF), the time scale of star formation and galactic wind must be
shorter than about 1 Gyr not to decrease the initial [$\alpha$/Fe]$>$0
(when $\alpha$-elements are mostly produced) to [$\alpha$/Fe]$\leq$0
(when iron is predominantly ejected). In other words, to reproduce
the observed trend of the [$\alpha$/Fe]-mass relationship, the total
duration of the star forming activity ought to scale with the galaxy
mass according to $\rm \Delta t_{SF} \propto M_{G}^{-1}$, which
opposes to the expectation from the CMR.

This point of contradiction cannot be avoided by the standard
chemo-spectro-photometric models -- both the closed-box ones by
\citet{Bressan94} and those with infall of primordial gas at early
epochs by \citet{Tantalo96,Tantalo98a} -- because they all stand on
the SDWG scheme and are tailored to reproduce the CMR. Therefore, they
fail to fit the companion mass-\alfa\ relationship. This issue has
been addressed by \citet{Tantalo01} who have carefully examined the
effect of key parameters of the classical models, namely the type and
efficiency of star formation and the role played by the rate of infall
on the degree of enhancement in \alfa. The main result of this
analysis is that the duration of star formation activity and age at
which galactic winds set in do not bear very much on the degree of
$\alpha$-enhancement. This in contrast is determined by the detailed
time dependence and intrinsic efficiency of star formation. In other
words, a galaxy whose stellar activity predominately occurs at very
early epochs (say within the first Gyr), even if followed by a long
tail of star formation at smaller efficiency (it does not matter
whether continuously or in recurrent episodes), would nowadays appear
as enhanced in $\alpha$-elements. Obviously, the opposite would hold
for a galaxy whose star formation is diluted at low, yet comparable
levels of intensity over long periods of time.

Finally, another short comment is worth here. Following
\citet{Ferrini00}, the above point of contradiction could stem from the
hypotheses that (i) SFR and consequent rate of supernova explosions
(mostly type II) are proportional to the gas content (and total galaxy
mass $M_G$ in turn); (ii) the binding energy of the gas is
proportional to $M_G$; and finally (iii) the star forming activity
stops after the galactic wind has occurred. All of the above hypotheses
are oversimplified descriptions of reality, for instance SF is known
to occur in molecular clouds, which are a subcomponent of the gas
content (in other words the SFR can be a non linear function of the
gas content), SF not necessarily halts after the first galactic wind
episode and finally the galactic winds can spread over significant
intervals of time. All this could destroy the simple relation between
the duration of SF and galaxy mass and perhaps alleviate the
discrepancy between the CMR and the $\alpha$-element problem
\citep{Ferrini00}.

The present approach based upon realistic dynamical models of early
type galaxies, in which neither star formation comes to an abrupt halt
nor galactic winds suddenly set up, see \citet{Chiocar01} for all
details, is perhaps a first step toward a self-consistent formulation
of the problem.

%%%%%%%%%Table 1  %%%%%
\begin{table*}  %%%[tH]
\begin{center}
\caption[]{Input parameters of the Tree-SPH models of EGs by \citet{Chiocar01}.}
\label{resume}
\setlength{\extrarowheight}{4pt}
\begin{tabular*}{133.5mm}{|l l l  l l c c c c| }
\hline
\multicolumn{1}{|c}{Model} &
\multicolumn{1}{c}{$\rm M_{T}$} &
\multicolumn{1}{c}{$\rm \langle \rho \rangle_{0}$} &
\multicolumn{1}{c}{$\rm \Delta T_{SF}$} &
\multicolumn{1}{c}{$\rm T_{PSF}$} &
\multicolumn{1}{c}{$\rm \langle \Psi\rangle$} &
\multicolumn{1}{c}{$\rm M_s$} &
\multicolumn{1}{c}{$\rm \sigma_B$} &
\multicolumn{1}{c|}{$\rm \%[{O\over Fe}]$} \\
\hline
\multicolumn{1}{|c}{} &
\multicolumn{1}{c}{$M_{\odot}$} &
\multicolumn{1}{c}{$M_{\odot}/kpc^{3}$} &
\multicolumn{1}{c}{$Gyr$} &
\multicolumn{1}{c}{$Gyr$} &
\multicolumn{1}{c}{$M_{\odot}/yr$} &
\multicolumn{1}{c}{$M_{\odot}$} &
\multicolumn{1}{c}{km s$^{-1}$} &
\multicolumn{1}{c|}{ } \\
\hline
\multicolumn{1}{|c}{(1)} &
\multicolumn{1}{c}{(2)} &
\multicolumn{1}{c}{(3)} &
\multicolumn{1}{c}{(4)} &
\multicolumn{1}{c}{(5)} &
\multicolumn{1}{c}{(6)} &
\multicolumn{1}{c}{(7)}&
\multicolumn{1}{c}{(8)}& 
\multicolumn{1}{c|}{(9)} \\
\hline
\multicolumn{9}{|c|}{Models  with  $\langle \rho \rangle_{0} \simeq 200\times \rho_{u}(z)$}\\
\hline
{{\bf 1A}} & $1\times 10^{13}$ & $4.2\times 10^{6}$ & 0.5 & 0.25 &3200 & $6.4\times 10^{11}$ & 530 &99\\ 
{{\bf 6A}} & $1\times 10^{9}$  & $4.2\times 10^{6}$ & 0.3 & 0.15 &0.07 & $3.2\times 10^{7}$ & 15 &99\\
\hline
\multicolumn{9}{|c|}{Models  with $\langle \rho \rangle_{0} \simeq  \rho_{u}(z)$}\\
\hline
{{\bf 1B}}    & $5\times 10^{13}$ & $3.8\times 10^{4}$ & 2.5 & 2.0 & 1760   & $4.4\times 10^{12}$ & 464 &99\\
{{\bf 2B}}    & $5\times 10^{12}$ & $3.8\times 10^{4}$ & 2.5 & 1.8 & 160    & $4.0\times 10^{11}$ & 210 &98\\
{{\bf 3B}}    & $1\times 10^{12}$ & $3.8\times 10^{4}$ & 2.5 & 3.0 & 25     & $6.4\times 10^{10}$ & 97  &98\\
{{\bf 5B}}    & $1\times 10^{10}$ & $3.8\times 10^{4}$ & 3.5 & 2.7 &0.11    & $4.0\times 10^{8}$  & 19  &94\\
{{\bf 6B-LD}} & $1\times 10^{9}$  & $5.5\times 10^{3}$ & 6.5 & 4.0 &0.002   & $1.3\times 10^{7}$  & 3   &44\\
{{\bf 6B}}    & $1\times 10^{9}$  & $3.8\times 10^{4}$ & 4.5 & 2.5 &0.005   & $2.5\times 10^{7}$  & 7   &76\\
{{\bf 6B-HD}} & $1\times 10^{9}$  & $5.8\times 10^{4}$ & 3.5 & 1.9 &0.022   & $7.7\times 10^{7}$  & 5   &91\\
{{\bf 7B}}    & $1\times 10^{8}$  & $3.8\times 10^{4}$ & 6.5 & 3.0 &0.0003  & $1.8\times 10^{6}$  & 4   &90\\
\hline
\end{tabular*}
\end{center}
\end{table*}

%=================== BEGIN SECTION ====================%
\section{Enhancement of $\alpha$-elements in dynamical models}

%=================== BEGIN SUBSECTION ====================%
\subsection{The dynamical models}

As we have already anticipated, \citet{Chiocar01} adopting the
monolithic scheme have calculated fully hydrodynamical models of
early-type galaxies with different total mass $\rm M_T$. The models
included dark and baryonic mass, $\rm M_D$ and $\rm M_B$ respectively,
in cosmological proportions ($\rm M_D=0.9\times M_T$ and $\rm
M_B=0.1\times M_T$), star formation, chemical evolution, energy
feed-back, and cooling by radiative processes.

In order to analyze the effect of the initial density of the
proto-galaxy three sets of models were considered: (i) a first group
(labeled {\bf A}) whose initial density is given by $\langle \rho
\rangle_{0} \simeq 200 \times \rho_{u}(z)$ (where $\rho_{u}(z)$ is the
mean density of the Universe as a function of red-shift); (ii) a
second group (labeled {\bf B}), in which the initial density has been
arbitrarily set equal to $\langle \rho \rangle_{0} \simeq
\rho_{u}(z)$; (iii) finally a third group limited to a unique value 
of the total mass ($\rm M_T=10^9\, M_{\odot}$) in which the initial
density has slightly varied above and below the case {\bf B}
value. They are hereinafter indicated by {\bf LD} (lower density) and
{\bf HD} (higher density).  For all other details the reader to refer
to \citet{Chiocar01}. The main results of this study can be summarized
as follows:

(i) Independently of the total mass, galaxies of high initial density
(case {\bf A}), undergo a prominent initial episode of star formation ever
since followed by quiescence.

(ii) The same applies to high mass galaxies of low initial density
(case {\bf B}), whereas the low mass ones undergo a series of burst-like
episodes that may stretch over a considerable fraction of their
lifetime. The details of their star formation history are very
sensitive to the value of the initial density as shown by the models
of the third group.
 
(iii) The mean and maximum metallicity increase with the galaxy mass.
Therefore these models can account for the CMR of EGs.
  
(iv) The model galaxies suffer galactic winds ejecting conspicuous
amounts of mildly enriched gas into the intra-cluster medium.
However, they do not follow the simple SDGW mechanism envisaged by
\citet{Larson74}: when the total thermal energy of the gas heated up
by supernova explosions (radiative cooling included) exceed the
gravitational potential energy of it, the remaining gas is suddenly
ejected and star formation is halted \citep[see for instance][\, for
more details on this type of models]{Bressan94,Tantalo96,Tantalo98a}.
In dynamical models the situation is more complicated. Looking at
radial velocity $\rm V_r$ of the gas and the escape velocity $\rm
V_{esc}$, at any time there is a layer a which some gas particles may
meet the condition $\rm V_r > V_{esc}$. All these gas particle can
freely leave the galaxy in form of galactic wind. Therefore galactic
winds are a kind of process taking place continuously, at least as
long as there are gas particles reaching and exceeding the above
condition. Furthermore, the total amount of gas lost in the wind $\rm
\Delta M_{gw}$ depends on $\rm M_T$. The following useful relations
are found $\rm \Delta M_{gw}/M_{B}=-0.09\, log M_T +1.256$ for models
B and $\rm \Delta M_{gw}/M_{B}=-0.06\, log M_T +0.99$ for models A.
{\it Low mass galaxies are somewhat more efficient emitters of partially processed material}.

(v) How does it happen that low mass galaxies may eject large
quantities of gas and yet form stars for long periods of time?  In a
normal or giant elliptical whose gravitational potential well is deep,
the gas tends to remain inside the potential well despite being heated
up by energy injection by the supernova explosions and stellar winds
accompanying star formation. A sort of equilibrium is reached in which
gas heating and cooling by radiative processes balance each other,
forcing gas to consumption by star formation. In contrast, in a low
mass galaxy, owing to the shallower gravitational potential well, the
same processes heat up gas, part of it spills over the potential well
and part flies to very low density regions so that star formation in
the inner regions declines. Eventually, the gas pushed away to large
distances but still bound to the galaxy cools down and falls back into
the central furnace thus re-fueling star formation. Therefore, this
latter keeps going for long periods of time in a series of burst-like
episodes till either gas is exhausted or what remains is heated up so
much (its gravitational binding energy in meantime gets smaller and
smaller compared to the thermal energy) that it eventually leave the
galaxy.

(vi) Finally, thanks to their star formation history these models have
the potential of improving upon one of the most embarrassing
difficulties encountered by the standard supernova driven galactic
wind (SDGW) model as far as the chemical abundances and abundance
ratios are concerned, i.e. the high $\rm [\alpha/Fe]$ ratio inferred
in massive EGs and the long duration of the star forming period
required by the SDGW model to explain the CMR.

In Table~\ref{resume} we summarize a few properties of the models by
\citet{Chiocar01} that are relevant to this study (we follow the same
identification coding used by \citet{Chiocar01} for the sake of an
easy comparison). Column (1) identifies the model. Column (2) gives
the total mass $\rm M_{T}$ in solar units. Column (3) gives the mean
initial density $\langle \rho \rangle_{0}$ in units of $\rm
M_{\odot}\,kpc^{-3}$. Columns (4) and (5) list the total duration of
the star formation activity $\rm \Delta T_{SF}$ in Gyr, and the age
$\rm T_{PSF}$ in Gyr at which the maximum stellar activity occurs,
respectively. Column (6) lists the mean star formation rate $\langle
\Psi \rangle$ in units of $\rm M_{\odot}/yr$. Column (7) gives the
present-day mass in stars in solar units. Column (8) contains the
central velocity dispersion, $\sigma$ of baryonic matter in $\rm km\,
s^{-1}$.

%=================== BEGIN SUBSECTION ====================%
\subsection{The detailed chemistry}

Although the description of the chemical enrichment adopted by
\citet{Chiocar01} was fully adequate to the purposes of N-body 
Tree-SPH simulations, the details of chemical abundances and abundance
ratios were out of reach owing to the numerical complexity of those
calculations.

To copy with this obvious limitation of the dynamical simulations, we
adopt here an indirect approach which albeit simple is sufficiently
accurate, i.e. we insert the SFR of dynamical models into the
chemical-evolution code of \citet{Portinari99}.

Since chemical evolution affects only the baryonic mass of the galaxy
it is sufficient to follow the evolution of an ideal galaxy having the
same baryonic mass $\rm M_B$, initially in form of gas. The chemical
models are in the so-called {\it closed-box} formulation
\citep[see][]{Tinsley80} as the dynamical ones. Finally, care has
been paid to secure that the yield of heavy elements and gas per
stellar generation are the same (or nearly the same) as those used in
the dynamical models so that the present-day metallicity and mass in
stars of the dynamical and chemical models with the same baryonic mass
and star formation history coincide.

{\it Sketch of the chemical models}. The equation governing the
abundances of elemental species in chemical models of galactic
evolution are well known so that no details are given here. The reader
is referred to \citet{Portinari98} for the detailed description of the
equations used also in the present study. Suffice here to recall a few
important ingredients:

(i) The equations are cast as function of the dimension-less variables

\begin{displaymath}
\rm 
G_g(t) = {M_g(t) \over M_B },  
   \,\,\,  G_s(t) = {M_s(t) \over  M_B }, 
\,\,\, 
\rm G_{g,i}(t)= G_g(t)\times X_i(t)
\label{mass_g}
\end{displaymath}

\noindent
where $\rm X_i(t)$ are the abundances by mass of the elemental species
$i$.\footnote{ Fifteen chemical elements are followed in detail,
namely $\rm H$, $\rm ^{4}He$, $\rm ^{12}C$, $\rm ^{13}C$, $\rm
^{14}N$, $\rm ^{16}O$, $\rm ^{20}Ne$, $\rm ^{24}Mg$, $\rm ^{28}Si$,
$\rm ^{32}S$, $^{40}Ca$, $^{56}Fe$, and the isotopic neutron-rich
elements $nr$ obtained by $\alpha$-capture on $\rm ^{14}N$, i.e. $\rm
^{18}O$, $\rm ^{22}Ne$, $\rm ^{25}Mg$.} By definition $\rm \sum_i
X_i(t)$=1.

(ii) The restitution fractions of the elements $i$ from a star of mass
$\rm M$ at the end of its life are obtained by means the matrix
formalism of \citep[see][]{Talbot71,Talbot73,Talbot75}.

(iii) No use of the instantaneous recycling is made (i.e. we take the
life time $\rm \tau_M $ of stars of mass $M$ into account.

(iv) We consider the separated contributions to nucleosynthetic ejecta
by single and binary stars. Single stars can terminate their life
quietly becoming white dwarfs and ejecting part of their mass in form
of stellar winds or exploding as Type II supernovae. Accreting white
dwarfs in binary stars are thought of to be the site of Type Ia
supernovae as long ago suggested by \citet{Greggio83} and
\citet{Mattgre86}. To evaluate the number of Type Ia supernovae we
need do fix the minimum ($\rm M_B$$_{m}$) and the maximum ($\rm
M_B$$_{M}$) value for the total mass ($\rm M_B$$_{M}$) of the binary
systems, the distribution function ($\rm f(\mu)$) of their mass
ratios, where $\rm \mu_{min}$ is the minimum value of it, and finally
the fraction $\Lambda$ of binaries with respect to the total. It is
assumed here that binary stars as a whole obey the same IMF of single
stars. We adopt $\rm B_{m}=3 M_{\odot}$, $\rm B_{M}=12 M_{\odot}$, and
$\Lambda=0.02$.

(v) The stellar ejecta are from \citet{Marigo96,Marigo97} and
\citet{Portinari98}, and incorporate the dependence on the initial
chemical composition as described by \citet{Portinari98}.

(vi) Finally, the stellar lifetimes $\rm \tau_M$ are from
\citet{Bertelli94}. We take also into account the dependence of $\rm
\tau_M$ on the initial chemical composition.

{\it The rate of star formation}. The rate of star formation (SFR) of
the dynamical models is according to the \citet{Schmidt1959} law
applied to each gas particle. Therefore, the total star formation rate
is also the Schmidt law applied to the whole system:

\begin{equation}
\rm  {d M_g(t) \over dt }=\nu \,M_g(t)^{\kappa}
\label{drho_dt}
\end{equation}

\noindent
where $\nu$=1 and $\kappa$=1 as in \citet{Chiocar01}. The SFR of the
dynamical models must be rescaled according to the normalization
adopted in the equations of the chemical models. Upon normalization,
the star formation rate becomes:

\begin{equation}
\rm \Psi(t)= {d\over dt} ({M_g \over M_B}) 
\label{sfr_nor}
\end{equation}

{\it The initial mass function.}  We have adopted here the same IMF
used by \citet{Chiocar01}, i.e. the \citet{Miller79} over the mass
interval $\rm M_l$=0.1 to $\rm M_u$=120 $M_{\odot}$. It is worth of
interest to look at the fractionary mass stored by the IMF in the mass
interval whose stars effectively contribute to nucleosynthesis in a
time interval shorter than the age of the galaxy.  A suitable choice
is the mass range $\rm M_n$=1$\, M_{\odot}$ to $\rm M_u$. The above
fraction $\zeta$ is

\begin{equation}
\rm \zeta =  \frac{\int_{M_n}^{M_u}\phi(M){\times}M{\times}dM}
{\int_{M_l}^{M_u}\phi(M){\times}M{\times}dM}
\label{zeta}
\end{equation}

\noindent
With the slope of the \citet{Miller79} IMF and the above mass limits,
$\zeta \simeq 0.25$.

%%%%%%%%%%%%%%%Figure 1 %%%%%%%%%%%%%%%
\begin{figure}
%\resizebox{\hsize}{!}{\includegraphics{fig_ps/OFe_Z_SFR.ps}}
\resizebox{\hsize}{!}{\includegraphics{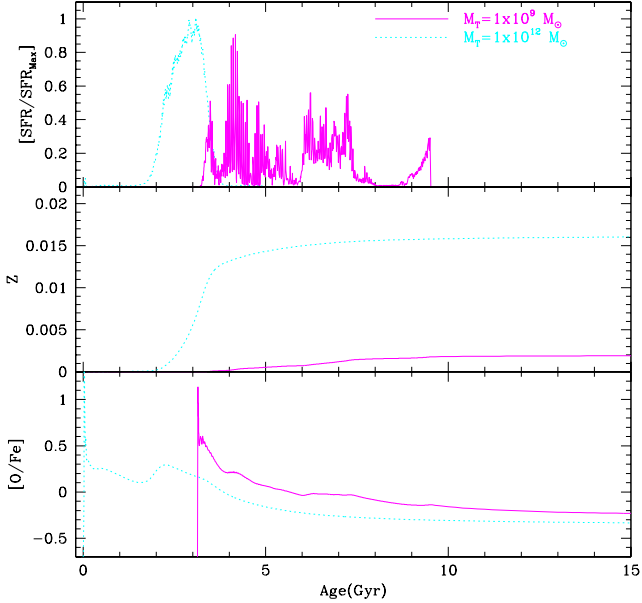}}
\caption{Bottom: the temporal variation of the abundance ratio
    [O/Fe] in models {\bf 3B} and {\bf 6B-LD}. Middle: the temporal
    variation of the metallicity in the same models. Top: the SFR of
    the same models normalized to the peak value SFR$_{\rm Max}$.}
\label{ratio_time}
\end{figure}

%%%%%%%%%%%%%%%Figure 2 %%%%%%%%%%%%%%%
\begin{figure}
%\resizebox{\hsize}{!}{\includegraphics{fig_ps/fig_m1e12_nolog.ps}}
\resizebox{\hsize}{!}{\includegraphics{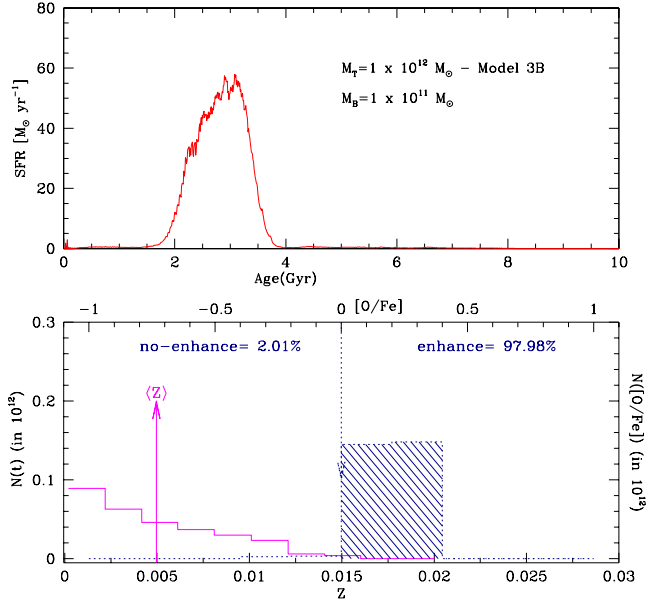}}
\caption{The top panel shows the SF history of the {\bf 3B} model  with
   total mass $\rm M_{T} = 10^{12} M_{\odot}$ and baryonic mass $\rm
   M_B = 10^{11} M_{\odot}$. The SFR is in $M_{\odot}$/yr and time is
   in Gyr. The bottom panel gives the cumulative number of living
   stars per bin of metallicity Z ({\it solid} line and bottom x-axis)
   and per bin of \OsFe\ ({\it dotted} line and top X-axis). The
   shaded zone shows the region populated by the stars with
   [O/Fe]$>$0, the $\alpha$-enhanced component. The percentage of
   star numbers with [O/Fe]$>$0 is about 98\%, whereas that with
   [O/Fe]$<$0 is 2 \%. }
\label{g1e12}
\end{figure}

%%%%%%%%%%%%%%%Figure 3 %%%%%%%%%%%%%%%
\begin{figure}
%\resizebox{\hsize}{!}{\includegraphics{fig_ps/fig_m1e9ld_nolog.ps}}
\resizebox{\hsize}{!}{\includegraphics{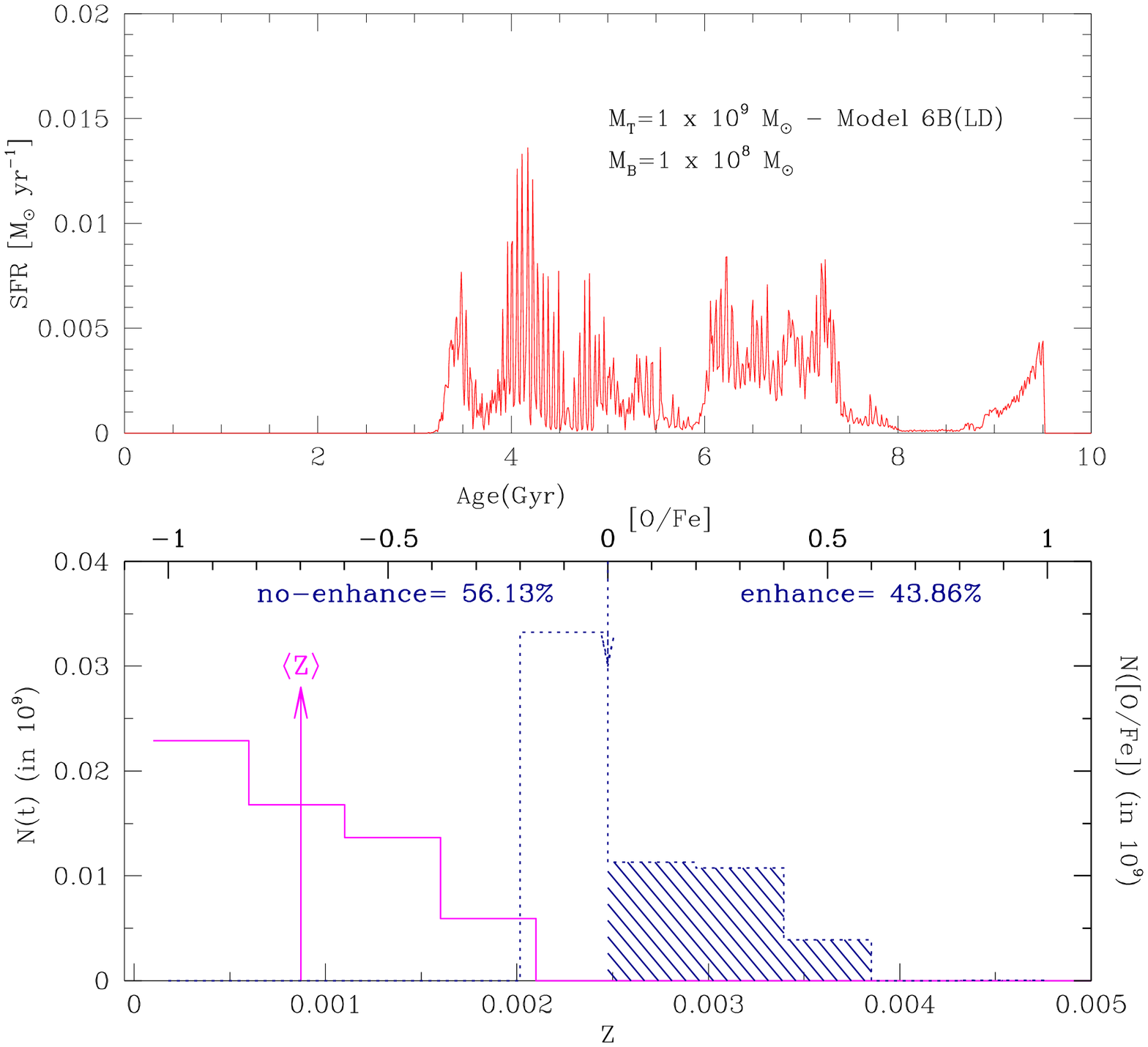}}
 \caption{The same of Fig.~\ref{g1e12} but for the {\bf 6B-LD} galaxy, the
 object with the lowest initial density and the longest duration of
 stellar activity. The total mass is $\rm M_{T} = 10^{9}M_{\odot}$,
 whereas the baryonic mass is $\rm M_{T} = 10^{9}M_{\odot}$. The
 percentage of $\alpha$-enhanced stars is now 44\%}
\label{g1e09}
\end{figure}

%=================== BEGIN SUBSECTION ====================%
\subsection{Chemical abundances and abundance ratios}

The chemical evolution of all the models presented in
Table~\ref{resume} has been followed and the results of interest here
are best represented by the series of Figs.~\ref{ratio_time} through
\ref{g1e09}.

In Fig.~\ref{ratio_time} we show the time variation of the SFR (top
panel), the metallicity Z (mid panel), and the abundance ratio [O/Fe]
(bottom panel) for the models {\bf 3B} with $\rm M_{T} = 10^{9}M_{\odot}$
and {\bf 6B-LD} with $\rm M_{T} = 10^{9}M_{\odot}$ as indicated. As
expected, during the period of stellar activity, the metallicity
increases, whereas when star formation is over the metallicity remains
either constant or slightly decreases by dilution with gas ejected by
old stars of low metal content. Looking at the time variation of the
abundance ratio [O/Fe] -- a measure of the $\alpha$-enhancement level
-- we note that

\noindent
(i) Within the first 0.1 Gyr the abundance ratios are lowered to
values comparable to the mean ones under the contamination by the
first Type Ia supernovae.

\noindent
(ii) Even a small amount of stellar activity changes the initial
abundance ratios to much lower values.

\noindent
(iii) Subsequent star formation does not alter the abundance ratios
significantly. See for instance the case of the $\rm 10^{12}\,
M_{\odot}$ galaxy in which the prominent burst occurring from 2 to 3.5
Gyr changes (increases) the [O/Fe] ratio by less than $\rm \Delta
[O/Fe]$=0.2 (less than a factor of two).

\noindent
(iv) Once star formation has ceased, almost immediately the effect of
Type II supernovae stops and that by Type Ia takes over first rapidly
and later slowly decreasing [O/Fe] towards the solar value or below
this.
  
In order to answer the question whether or not the present-day stellar
content of these galaxies would appear to be enhanced in
$\alpha$-elements, we look at the cumulative distribution of the
number of living stars in different bins of [O/Fe]. From an
operational point of view a galaxy will be considered dominated by an
$\alpha$-enhanced stellar content if $\rm \sum N([O/Fe]>0)/N_T > 0.5$,
where $\rm \sum N([O/Fe])$ is the number of stars in bins with
[O/Fe]$>$0 and $\rm N_T$ is their total number. The opposite if $\rm
\sum N([O/Fe]<0)/N_T < 0.5$. The situation for the two galaxies under
examination is shown in Fig.~\ref{g1e12} ($\rm 10^{12}\, M_{\odot}$)
and Fig.~\ref{g1e09} ($\rm 10^{9}\, M_{\odot}$).

The massive galaxy (Fig.~\ref{g1e12}), characterized by a rather short
($\sim 2.5$ Gyr) and prominent burst of star formation activity has
about 98\% of its stellar content in bins with with \OsFe$>0$. Whereas
the low mass, low initial density galaxy (Fig.~\ref{g1e09}), whose star
formation stretched over long periods of time in several recurrent
episodes, has the majority of stars in bins with
\OsFe$< 0$ ($\sim$44\%). 

The percentage of stars with $\rm N([O/Fe])/N_T > 0.5$ are listed in
column (9) of Table~\ref{resume}. Finally, in columns (3) and (4) and
of Table~\ref{index} to better described below we list the maximum,
$\rm Z_{Max}$, and the mean metallicity, $\rm \langle Z \rangle$, of
our model galaxies.

%%%%%%%%%%%%%%%Figure 4 %%%%%%%%%%%%%%%
\begin{figure}
%\resizebox{\hsize}{!}{\includegraphics{fig_ps/shape_SFR.ps}}
\resizebox{\hsize}{!}{\includegraphics{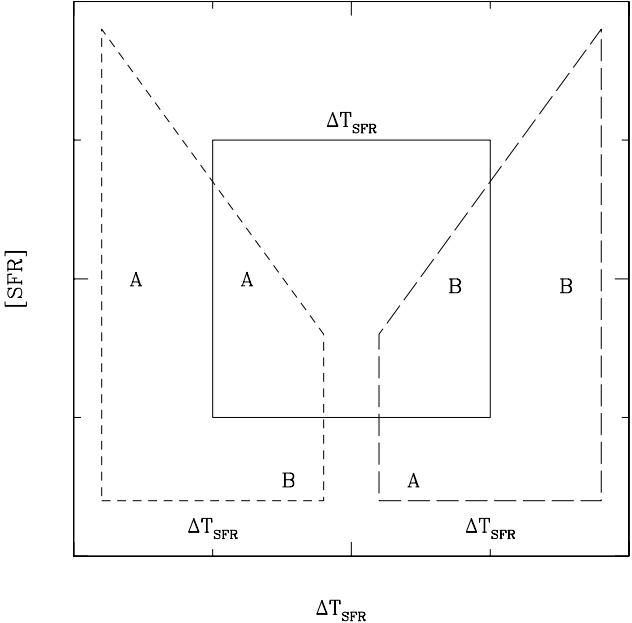}}
 \caption{Schematic SFRs of our simulations. See the text for more
 details.}
\label{schem-sfr}
\end{figure}

%%%%%%%%%%%%%%%Figure 5 %%%%%%%%%%%%%%%
\begin{figure}
%\resizebox{\hsize}{!}{\includegraphics{fig_ps/Perc_enhance.ps}}
\resizebox{\hsize}{!}{\includegraphics{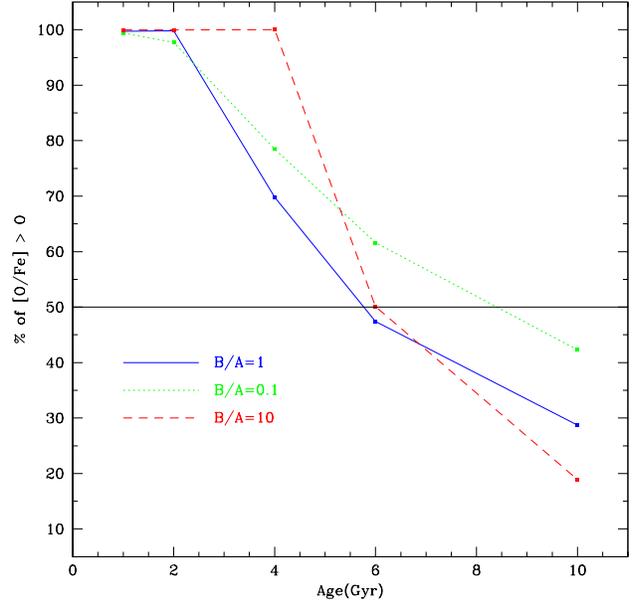}}
 \caption{Degree of enhancement in models with different shape and
 duration of the SFR. The dotted line is for B/A=0.1, the solid line
 for B/A=1, and the dashed line for B/A=10. See the text for more
 details}
\label{degree-enh}
\end{figure}

%=================== BEGIN SECTION ====================%
\section{What does determine the enhancement of $\alpha$-elements?}

The entries of Table~\ref{resume} show that the percentage of stars
with [O/Fe]$>$0 decreases with the total duration of the star
formation activity $\rm \Delta T_{SF}$. However all models but for the
case {\bf 6B-LD} have percentages of $\alpha$-enhanced stars higher than
50\%. In order to understand under which circumstances (shape of the
SFR and $\rm \Delta T_{SF}$) a galaxy would build a stellar population
with a percentage of $\alpha$-enhanced stars less than 50\%, we have
performed the following experiment. We have taken one of the most
unfavorable cases, i.e. a typical galaxy with $\rm M_T=10^{12}\,
M_{\odot}$ and hence $\rm M_B=10^{11}\, M_{\odot}$, imposed that the
same amount of mass in stars is build up at the end of the star
forming activity, and varied both the shape and duration, $\rm \Delta
T_{SF}$ of the SFR.

We adopt a simple analytical expression for the SFR as illustrated in
Fig.~\ref{schem-sfr}. This is idealized to a trapezium of sides A, B
and $\rm \Delta T_{SF}$. By varying the ratio B/A, the shape of the
SFR changes from case (1) in Fig.~\ref{schem-sfr} with B/A$ <<$ 1 (SFR
peaked in the past), to a constant SFR with B/A=1 of case (2) in
Fig.~\ref{schem-sfr}, and finally to case (3) in Fig.~\ref{schem-sfr}
(SFR peaked toward the present). The total area of the trapezium
(rectangle) yields the total mass $\rm M_s$ in stars build up in the
galaxy. The galaxy under consideration has a total mass in stars of
about $\rm 6.4\times 10^{10}\, M_{\odot}$.

The following relationship holds between $\rm M_s$ and SFR

\begin{equation}
  \rm   M_s = 0.5 \times A \big[ 1 + {B\over A}  \big]
  \times \Delta T_{SF}
\end{equation}

\noindent
from which we may easily derive an analytical expression for the SFR
to be plugged into the chemical code. We have explored the following
combinations of the parameters B/A=0.1, 1.0, and 10 and $\rm \Delta
T_{SF}$=1, 2, 4, 6 and 10 Gyr. The results are displayed in
Fig.~\ref{degree-enh}, which plots the present-day percentage of
$\alpha$-enhanced stars as a function of $\rm \Delta T_{SF}$ and the
ratio B/A (the dotted line for B/A=0.1, the solid line for B/A=1, and
the dashed line for B/A=10). It is soon evident that $\rm \Delta
T_{SF}$ is the dominant parameter, whereas the ratio B/A plays a
marginal role. With the prescription for Type Ia supernovae we have
adopted, to fall below the border line of 50\%, $\rm \Delta T_{SF}$
should be longer than about 6 Gyr and the stellar activity should be
uniformly diluted. This is indeed the case of 6B-LD model which finds
confirmation by the present analysis. A counter-example is the case of
the 7B model with $\rm M_T=10^8 \, M_{\odot}$ whose $\rm \Delta
T_{SF}\simeq$6.5 Gyr but the SFR is mostly confined between 2 to 4 Gyr
with the peak at about 3 Gyr. The reason for that is the initial
density which is slightly higher than that of case {\bf 6B-LD} (see
Table~\ref{resume} and \citet{Chiocar01} for all details). The total
duration of the SFR and not its shape drives the level of enhancement.

%=================== BEGIN SECTION ====================%
\section{Mean narrow band indices for galaxies }

In this section we evaluate the narrow band indices for our model
galaxies in order to check how far they can go in reproducing
important relationships such as the $\rm Mg_2$ versus $\sigma$
\citep{Burstein88} and the $\rm H_{\beta}$ versus $\rm Mg_2$ and/or $\rm MgFe$,  
see for instance \citet{Gonzalez93,Trager20a,Trager20b}.

In presence of enhancement in \alfa\ one has to modify
the relationship between the total metallicity $Z$ and the iron content
\FeH\ by suitably defining the enhancement factor $\Gamma$. For more 
details see \citet{Tantalo98b} and \citet{Tantalo02}.

Let us split the metallicity $Z$ in the sum of two terms

\begin{equation}
\rm  Z = \sum_j X_j + X_{Fe}
\label{sum1}
\end{equation}

\noindent
where $\rm X_j$ are the abundances by mass of all heavy elements but
Fe, and $\rm X_{Fe}$ is the same for Fe. Recasting eq.(\ref{sum1}) as

\begin{equation}
\rm Z = \frac{X_{Fe}}{X_H} X_H \left[ 1+ \frac{\sum_j X_j}{X_{Fe}} \right]
\label{sum2}
\end{equation}

\noindent
and normalizing it to the solar values we get

\begin{equation}
\rm 
\left(\frac{Z}{Z_{\odot}}\right)= 
    \left(\frac{X_{Fe}}{X_H} \right) 
       \left(\frac{X_H}{X_{Fe}} \right)_{\odot}
          \left(\frac{X_H}{X_{H,\odot}} \right) 
   \frac{\left[1+ \frac{\sum_j X_j}{X_{Fe}}\right]} 
          {\left[1+ \frac{\sum_j X_j}{X_{Fe}}\right]_{\odot}} 
\label{sum3}
\end{equation}

\noindent 
from which we finally obtain  

\begin{equation}
\rm \left[ \frac{Fe}{H} \right] = \lg{\left(\frac{Z}{\Zsun}\right)} -
\lg{\left(\frac{X}{\Xsun}\right)} - \Gamma
\label{feh1}
\end{equation}

\noindent
which provides also the definition of $\Gamma$. Obviously for solar
metallicity $\rm Z_{\odot}$ and solar-scaled mixture ($\Gamma = 0.0$)
eq.(\ref{feh1}) yields \FeH $=0.0$. In general, when dealing with
enhanced chemical mixtures, it suffices to determine $\Gamma$
according to the degree of enhancement and simply re-scale the
zero-point for the iron content of the Sun, i.e \FeH=-$\Gamma$). {\it
Remarkably, enhancing \alfa\ simply reduces to decreasing \FeH}.

\noindent
Table~\ref{enh-deg} lists the initial chemical composition and the
iron abundances corresponding to $\Gamma$=0 and $\Gamma$=0.356, the
same as the stellar models and isochrones of \citet{Salasnich20} on
which our indices are ultimately based.

%%%%%%%%%Table 2  
\begin{table}
\small
\begin{center}
\caption[]{Real \FeH\ for solar-scaled and \alfe\ isochrones as a 
function of the initial chemical composition [X,Y,Z].}
\label{enh-deg}
\begin{tabular}{|c c c | r| r|}
\hline
\multicolumn{1}{|c}{} &
\multicolumn{1}{c}{} &
\multicolumn{1}{c|}{} &
\multicolumn{1}{c|}{$\Gamma=0.0$} &
\multicolumn{1}{c|}{$\Gamma=0.356$} \\
\hline 
\multicolumn{1}{|c}{Z} &
\multicolumn{1}{c}{Y} &
\multicolumn{1}{c|}{X} &
\multicolumn{1}{c|}{[Fe/H]} &
\multicolumn{1}{c|}{[Fe/H]} \\
\hline
 0.008& 0.248 & 0.7440 & --0.3972 & --0.7529 \\
 0.019& 0.273 & 0.7080 &   0.0000 & --0.3557 \\
 0.040& 0.320 & 0.6400 &   0.3672 &   0.0115 \\
 0.070& 0.338 & 0.5430 &   0.6824 &   0.3267 \\
\hline
\end{tabular}
\end{center}
\end{table}

Once $\Gamma$ is known for the stellar population under consideration,
to proceed further we need a calibration expressing the indices as a
function of it. To this purpose we make use of the calibration by
\citet{Borges95} for \mgii\  in which the effect of enhancement is 
considered. According to \citet{Borges95}, the index \mgii\ of a
``star'' with enhanced content of Mg is given by the relation

%%%%%%%%%%%%%Equation 
\begin{displaymath}
\rm  {\ln}{Mg_{2}} = -9.037 + 5.795 \frac{5040}{T_{eff}} + 0.398 \log{g}
+ ~~~~~~~~
\end{displaymath}
\begin{equation}
\rm  ~~~~~~~~0.389 \left[ \frac{Fe}{H} \right] - 0.16 \left[ \frac{Fe}{H}
\right]^{2} + 0.981 \left[ \frac{Mg}{Fe} \right] 
\label{mg2}
\end{equation}

\noindent
which holds for effective temperatures and gravities in the ranges
$\rm 3800 < T_{eff} < 6500$ K and $\rm 0.7 < \log{g} < 4.5$. To make
use of eq.(\ref{mg2}), we need \MgsFe\ as function of $\Gamma$. The
correspondence between \MgsFe\ and $\Gamma$ can be obtained in the
following way: from the pattern of abundances by
\citet{Anders89,Grevesse91} and \citet{Grevesse93} we get the
approximated relation

%%%%%%%%%%%%%Equation 
\begin{equation}
\rm \Gamma \simeq   0.8\left[\frac{Mg}{Fe}\right] 
\label{mgfe}
\end{equation}

\noindent
According to which \MgsFe=0.0 for $\Gamma$=0 (solar-scaled
abundances), and \MgsFe=0.444 for $\Gamma$=0.356.

For all remaining indices we use the \citet{Worthey94} fitting
functions in which the enhancement of \alfa\ is taken into account
only via the re-scaled value of \FeH\ of eq.(\ref{feh1}).

%%%%%%%%%Table 3 (Analytical relationships for indices ) %%%%%
\begin{table*}  %%%%%%%%%%%%%[tH]
\begin{center}
\caption[]{Coefficients for the relation (\ref{secord}) and the calibration by \citet{Borges95}.}
\label{coeff}
\begin{tabular*}{148mm}{|c| r r r r r r|}
\hline
\multicolumn{1}{|c|}{$I_{i}$} &
\multicolumn{1}{c}{$a_{1,i}$} &
\multicolumn{1}{c}{$a_{0,i}$} &
\multicolumn{1}{c}{$b_{1,i}$} &
\multicolumn{1}{c}{$b_{0,i}$} &
\multicolumn{1}{c}{$c_{0,i}$} &
\multicolumn{1}{c|}{{$\delta_i$}} \\
\hline
$\rm H_{\beta}$ &--0.565 $\pm$ 0.134 &--1.356 $\pm$ 0.228 & --0.295 $\pm$ 0.140 &--0.673 $\pm$ 0.029 &  0.205 $\pm$ 0.015 & 2.89 \\
$\rm Mg_{2}$    &  0.089 $\pm$ 0.034 &  0.135 $\pm$ 0.056 & --0.072 $\pm$ 0.027 &  0.165 $\pm$ 0.030 &--0.093 $\pm$ 0.010 & 0.10 \\
$\rm \langle Fe \rangle$ & 0.488 $\pm$ 0.132 & 0.845 $\pm$ 0.580 & 0.357 $\pm$ 0.113 & 1.725 $\pm$ 0.185 & 0.898 $\pm$ 0.026 & 2.18 \\
$\rm Mg_{b}$    &  1.103 $\pm$ 0.234 &  1.676 $\pm$ 0.675 &   0.115 $\pm$ 0.298 &  2.011 $\pm$ 0.359 &  0.806 $\pm$ 0.066 & 2.09 \\
\hline
\end{tabular*}
\end{center}
\end{table*}

Looking at the response of indices to variations of the age, $\rm
\log(t)$ in Gyr, metallicity $\rm \log(Z/Z_\odot)$, and degree of
enhancement $\Gamma$ of SSPs, we derive the following analytical
relationships,

\begin{equation}
	\begin{array}{ll}
\rm I_{i} = \left( a_{1,i} \Gamma + a_{0,i} \right) \log(t)\\
~~~~~~+\left( b_{1,i} \Gamma + b_{0,i} \right)
\log{\left(\frac{Z}{Z_{\odot}}\right)}+ c_{0,i} \Gamma + 
   \delta_{i} 
	\end{array}
\label{secord}
\end{equation}

\noindent
where $\rm I_{i}$ for $i$=1 to 4 stand for the \Hbeta, \mgii, \MFe\
and \mgb, respectively. The coefficients are tabulated in
Table~\ref{coeff} for the \citet{Borges95} calibration. These
analytical fits are accurate enough for the purposes of the present
study.

Given these premises, we derive the indices for our model galaxies as
follows:

\noindent
(i) With the aid of the cumulative distributions of living stars per
[O/Fe] bins (see the bottom panels of Figs.~\ref{g1e09} and
\ref{g1e12} and those for all remaining models not shown here for the
sake of brevity), we estimate the mean value of [O/Fe] weighed on the
number of stars per bin and consider it as an estimate of the
enhancement factor $\Gamma$. This is found to weakly correlate with
the velocity dispersion of the galaxy $\Gamma= 0.072\,log \sigma +
0.134$.

\noindent
(ii) From the history of star formation (top panels of the same
Figures.) we estimate the mean age $\rm T_{BS}$ of the bulk stars. To
do so one has to convert the rest-frame ages of the galaxy models into
an absolute age by means of a cosmological model of the Universe. To
make things simple we adopt here the Freedman model with $\rm H_0$=50
km s$^{-1}$ Mpc$^{-1}$, $\rm q_0$=0 and red-shift of galaxy formation
$\rm z_{for}=5$, which yield the maximum age for galaxies $\rm
T_G$=16.5 Gyr.  The age $\rm T_{BS}$ is therefore $\rm
T_{BS}=T_G-T_{PSF}$, where $\rm T_{PSF}$ is the time at which the peak
of star formation occurred. These quantities are listed in columns
(5) and (6) of Table~\ref{index}.

\noindent
(iii) Finally, we make use of relations (\ref{secord}) to derive the
indices $\rm H_{\beta}$, $\rm Mg_2$, $\rm \langle Fe \rangle$, and
$\rm Mg_b$ for both the maximum and mean metallicity (age $\rm T_{BS}$
and $\Gamma$ are the same). The results are listed in Table~\ref{index}.

%%%%%%%%%Table 4 (indices versus sigma relationship ) %%%%%
\begin{table*}  %%%%%%%%%%%%%[tH]
\begin{center}
\caption[]{Theoretical indices for all the models under considerations.
 The calibration in use is by \citet{Borges95}.}
\label{index}
\begin{tabular*}{155mm}{|l l r r c r| r r r r| r r r r|}
\hline
\multicolumn{1}{|c}{Model} &
\multicolumn{1}{c}{$\rm M_T$} &
\multicolumn{1}{c}{$\rm Z_{Max}$} &
\multicolumn{1}{c}{$\rm \langle Z\rangle$} &
\multicolumn{1}{c}{$\Gamma$} &
\multicolumn{1}{c|}{$\rm T_{BS}$} &
\multicolumn{1}{c}{$\rm H_{\beta}$} &
\multicolumn{1}{c}{$\rm Mg_2$} &
\multicolumn{1}{c}{$\rm \langle Fe\rangle$} &
\multicolumn{1}{c|}{$\rm Mg_b$}  &
\multicolumn{1}{c}{$\rm H_{\beta}$} &
\multicolumn{1}{c}{$\rm Mg_2$} &
\multicolumn{1}{c}{$\rm \langle Fe\rangle$} &
\multicolumn{1}{c|}{$\rm Mg_b$} \\
\hline
\multicolumn{6}{|c|}{}&\multicolumn{4}{c|}{Maximum Z}&\multicolumn{4}{c|}{Mean Z}\\
\hline
1A      & $1\times 10^{13}$ & 0.020  & 0.005  & 0.53 & 16.3 & 1.47 & 0.26 & 2.49 & 3.09 & 1.79 & 0.14 & 1.56 & 1.91\\
6A      & $1\times 10^{9} $ & 0.004  & 0.002  & 0.66 & 16.3 & 1.89 & 0.11 & 1.18 & 1.33 & 2.08 & 0.02 & 0.57 & 0.54\\
\hline
1B      & $5\times 10^{13}$ & 0.025  & 0.008  & 0.14 & 14.5 & 1.30 & 0.28 & 3.15 & 3.98 & 1.61 & 0.19 & 2.33 & 2.99\\
2B      & $5\times 10^{12}$ & 0.018  & 0.007  & 0.27 & 13.0 & 1.51 & 0.24 & 2.69 & 3.36 & 1.75 & 0.17 & 2.02 & 2.55\\
3B      & $1\times 10^{12}$ & 0.015  & 0.005  & 0.19 & 12.0 & 1.57 & 0.23 & 2.65 & 3.31 & 1.86 & 0.14 & 1.86 & 2.36\\
5B      & $1\times 10^{10}$ & 0.006  & 0.003  & 0.15 & 11.0 & 1.85 & 0.15 & 2.01 & 2.54 & 2.03 & 0.10 & 1.53 & 1.97\\
6B-LD   & $1\times 10^{9} $ & 0.001  & 0.002  & 0.05 &  6.0 & 2.69 & 0.01 & 0.59 & 0.75 & 2.47 & 0.04 & 1.14 & 1.39\\
6B      & $1\times 10^{9} $ & 0.004  & 0.002  & 0.12 & 10.0 & 2.01 & 0.12 & 1.72 & 2.18 & 2.14 & 0.08 & 1.38 & 1.78\\
6B-HD   & $1\times 10^{9} $ & 0.006  & 0.002  & 0.19 & 12.0 & 1.81 & 0.15 & 1.99 & 2.52 & 2.11 & 0.07 & 1.20 & 1.57\\
7B      & $1\times 10^{8} $ & 0.003  & 0.001  & 0.18 &  8.0 & 2.22 & 0.08 & 1.37 & 1.68 & 2.51 & 0.01 & 0.58 & 0.73\\
\hline
\end{tabular*}
\end{center}
\end{table*}

%%%%%%%%%%%%%%%Figure 6 %%%%%%%%%%%%%%%
\begin{figure}
%\resizebox{\hsize}{!}{\includegraphics{fig_ps/Mg2_sigma.ps}}
\resizebox{\hsize}{!}{\includegraphics{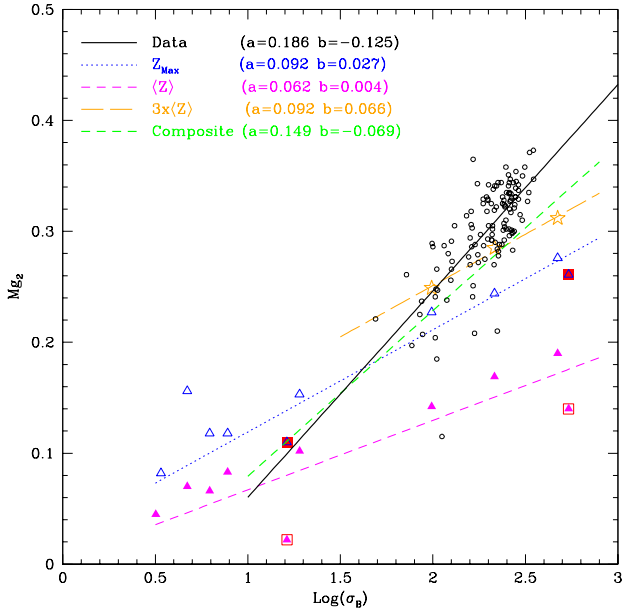}}
 \caption{The $\rm Mg_2$ versus velocity dispersion $\sigma_B$.  The
 small open circles are the observational data from
 \citet{Trager20a,Trager20b} and the solid line is their least squares
 fit. The open and filled triangles are the theoretical models of type
 {\bf B} listed in Table~\ref{index} for the maximum and mean metallicity,
 respectively. Their least square fits are shown by the dashed ($\rm
 Z_{Max}$) and dotted ($\rm \langle Z \rangle$) lines. The three empty
 stars show the same, however limited to the three most massive
 galaxies, but for a metallicity arbitrarily increased to $\rm 3\times
 \langle Z \rangle$. The thick dashed line shows the composite case in
 which the metallicity of the three most massive galaxies is $\rm
 3\times \langle Z \rangle$ and that of the less massive ones is $\rm
 \langle Z \rangle$. The filled squares ($\rm Z_{Max}$) and
 boxed-filled triangles ($\rm langle Z \rangle$ show the case of
 models {\bf A}.}
\label{mg2_sigma}
\end{figure}

{\it The $\rm Mg_2$ vs $\sigma$ relation.} This is shown in
Fig.~\ref{mg2_sigma} and compared with the observational data by
\citet{Trager20a,Trager20b} for a sample of local EGs. The least
squares fit of the observational data yields the relation $\rm
Mg_2=0.186\,log\sigma-0.125$ (where $\sigma$ is in km\,s$^{-1}$). The
theoretical results are plotted separately for each of the
metallicities in use: the filled triangles are for $\rm \langle Z
\rangle$ (a more realistic approximation of reality) and the open
triangles are for $\rm Z_{Max}$. The least squares fit of the data
yields $\rm Mg_2=0.063\,log\sigma+0.001$ for $\langle Z \rangle$ and
$\rm Mg_2=0.094\,log\sigma+0.063$ for $Z_{Max}$. The slope is about
half of what observed. Where is the source of disagreement? There are
several candidates: (i) the velocity dispersion, (ii) the calibration
in use, (iii) the set of parameters $\rm T_{BS}$, $\rm Z$ and $\Gamma$
assigned to the model galaxies. Limiting the discussion to the range
of massive models for which data exist, we note:

\noindent
(i) The velocity dispersion is not likely because it should be
decreased by a factor of two and in any case the relation would simply
horizontally shift without changing its slope.

\noindent
(ii) The calibration is more uncertain, even though the one in use
here allowing for the maximum effect. In this context one should,
however, keep in mind that, contrary to what one may guess, increasing
$\Gamma$ (higher enhancement) does not imply a stronger $\rm Mg_2$
because of the counter-reaction by [Fe/H].

\noindent
(iii) Assuming that our velocity dispersions and calibration are
correct, inspection of the weight of $T_{BS}$, $\Gamma$, and $\rm Z$
on building up the final result reveals that the metallicity
dominates. The three empty stars show the effect of arbitrarily
increasing $\rm \langle Z \rangle$ by a factor of 3, limited to the
three most massive galaxies. This implies that the $\rm Mg_2$ vs
$\sigma$ relation reflects the mass-metallicity relation, rather a
sequence in the enhancement level.

Are the mean (and maximum) metallicity reasonable? Is there any
plausible effect making a massive galaxy more efficient in building
metals? With the IMF adopted here and in the dynamical models, the
production of metals is scarcely efficient because of the low $\zeta$,
Increasing this for all galaxies would simply shift vertically the
$\rm Mg_2$ vs $\sigma$ relation. However, if $\rm \zeta$ increases
with the galaxy mass (top-heavy IMF), theory and data could be
reconciled. Although this possibility has often been advanced by
several authors in different astrophysical contexts -- see the
arguments given by \citet{Chiosi98} and \citet{Larson98} -- further
exploring the subject is beyond the scope of this paper. Suffice to
conclude that dynamical models almost match the observational $\rm
Mg_2$ vs $\sigma$ relation provided some fine tuning of the chemical
yields per stellar generation is applied. Just for the sake of
curiosity, we derive the relationship one would obtain by letting the
mean metallicity to increase from the present values to $\rm 3\times
\langle Z \rangle$ as the galaxy mass mass increases from $10^9$ to
$\rm 5\times 10^{12}\, M_{\odot}$. We get $\rm
Mg_2=0.149\,log\sigma-0.070$, which is about right.

%%%%%%%%%%%%%%%Figure 7 %%%%%%%%%%%%%%%
\begin{figure}
%\resizebox{\hsize}{!}{\includegraphics{fig_ps/Hb_Mg2_MgFe.ps}}
\resizebox{\hsize}{!}{\includegraphics{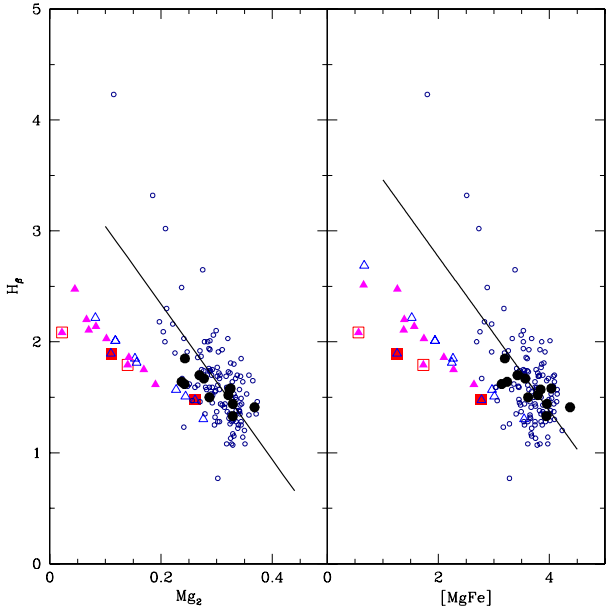}}
\caption{The $H_{\beta}$ versus $Mg_2$ and $MgFe$ planes. The meaning of
  the symbols is the same as in Fig.~\ref{mg2_sigma}. We have
  also added the small sample of EGs in Fornax by \citet{Kuntschner00}
  (big filled circles). The line is the linear fit of all
  observational data.}
\label{hbeta_mg2_mgfe}
\end{figure}

{\it The $\rm H_{\beta}$ versus $\rm Mg_2$ and $\rm MgFe$ planes.}
The index-index planes are customarily used to infer age and
metallicity of early-type galaxies \citep{Gonzalez93,Bressan96,
Trager20a, Trager20b, Kuntschner00, Kuntschner01}. In
Fig.~\ref{hbeta_mg2_mgfe} we present the $\rm H_{\beta}$ vs $\rm Mg_2$
and $\rm H_{\beta}$ vs $\rm MgFe$ planes left and right panels
respectively, for our models, single stellar populations of similar
metallicity, and observational data for a small sample of EGs in the
Fornax cluster by \citet{Kuntschner00} and the wider sample of nearby
galaxies by \citet{Trager20a}. The following remarks can be made:

\noindent
(i) Our models lie on a strip which roughly correspond to that covered
by SSP of different metallicity and age in the range say 15 to 8-10
Gyr. This expected because all models have terminated their star
formation far in the past. Even the most extreme case has ceased star
formation 6 Gyr ago.

\noindent
(ii) Also in these planes we notice the same kind of difficulty
encountered with the $\rm Mg_2$ vs $\sigma $ relation, i.e. our model
galaxies do not make enough metals so that even the most massive ones
only marginally reach the region crowed by the data. The same kind of
arguments brought about to cure the $\rm Mg_2$ vs $\sigma$ relation
apply here. Increasing the metal content of the model galaxies is
less of a problem.

%=================== BEGIN SECTION ====================%
\section{Concluding remarks}

In this paper we have derived the present-day level of enhancement in
\alfa\ in the stellar content of the N-Body Tree-SPH models of
EGs by \citet{Chiocar01}, cast light on the main cause of it, derived
the mean spectral indices of the model galaxies, and compared them to
the observations with particular attention to the $\rm Mg_2$-$\sigma$
relation and the $\rm H_{\beta}$ - $\rm Mg_2$--[MgFe] planes.

Combining the results of \citet{Chiocar01} summarized in Section 3 and
those obtained here the following remarks can be made:

(i) The star formation history of EGs depends on their total mass and
initial density, i.e. the physical conditions of the environment in
which they have formed. High-density systems, owing to their very
early and short star forming activity would now appear as old
galaxies, whose stellar content is enhanced in \alfa. Low-density
systems of high mass would resemble the previous ones, their stellar
content should, however, span a broader range of ages. Most of the
component stars should be enhanced in \alfa\ but with a broader
distribution of abundance ratios [$\alpha$/Fe]. At decreasing mass the
star formation activity should become more diluted in time, the
stellar populations should span wider and wider age ranges, and the
level of enhancement should progressively tend to solar.  The extreme
case is reached with the low-mass, low-density systems which could
undergo many episodes of star formation up to very recent times, thus
resembling young objects with solar or sub-solar partitions of
elements. It is worth recalling here that $\Gamma$ or equivalently
[O/Fe] is found to weakly increase with the velocity dispersion (mass)
of the galaxy.

(ii) The scenario sketched above is perhaps confirmed by the many
observational studies based on different kinds of diagnostic. For
instance \citet{Kuntschner98a,Kuntschner00,Kuntschner01} studying
early-type galaxies in the Fornax cluster find the EGs to be coeval
and to have metallicities varying from solar to three times solar, and
the ratio [Mg/Fe] --inferred from the indices $\rm Mg_2$ and Fe3-- to
weakly correlate with the velocity dispersion. In contrast the data
for nearby field galaxies by \citet{Gonzalez93,Bressan96}, and
\citet{Trager20a,Trager20b}, seem to indicate a much broad range of
ages as perhaps compatible with our model galaxies of low initial
density. Finally, we call attention on the recent result by
\citet{Poggianti01} for dwarf galaxies confirming the suggestion made
by \citet{Bressan96} and \citet{Trager20a,Trager20b} of a recent star
formation in these systems.

(iii) The $\rm Mg_2$ vs $\sigma$ relation mainly reflects the
metallicity-$\sigma$-mass sequence rather than enhancement in
\alfa\ and/or age.

(iv) Although the dynamical models and companion chemical models we
have discussed do not fully match the high metallicities inferred from
other diagnostics and some improvement is required, they however seem
to offer a coherent scenario accounting for many observed properties
of EGs, thus perhaps indicating that we are on the right track.
 
%=============== ACKNOWLEDGMENTS ======================%
\begin{acknowledgements} We like to  thank Dr. F. Ferrini for several
constructive suggestions in his referee report. C.C. likes to
acknowledge the friendly hospitality and stimulating environment
provided by ESO in Garching where this paper has been written during
leave of absence from the Astronomy Department of the Padua
University. This study has been financed by the Italian Ministry of
Education, University, and Research (MIUR).
\end{acknowledgements}

%=================== BIBLIOGRAPHY ======================%
\bibliographystyle{apj}           % Files .bst
\bibliography{mnemonic,biblio}    % Files .bib
%=================== END   DOCUMENT ===================%

\end{document}